\def\gsim{\mathrel{\scriptstyle{\buildrel > \over \sim}}}

\magnification=1200
\baselineskip=17pt

\vskip 32pt
\centerline {\bf METALLIC SURFACE RECONSTRUCTION DRIVEN}
\centerline {\bf BY FRUSTRATED ANTIFERROMAGNETISM}
\vskip 50pt
\centerline{J. P. Rodriguez\footnote {$^*$} 
{Permanent address: Dept. of Physics
and Astronomy, California State University, 
Los Angeles, CA 90032, USA.}
and Emilio Artacho}
\medskip
\centerline{\it Departamento de F\'{\i}sica de la Materia Condensada,
C-III, and Instituto Nicol\'as Cabrera,}
\centerline{\it Universidad Aut\'onoma de Madrid,
Cantoblanco, 28049 Madrid, Spain.}
\vskip 30pt
\centerline  {\bf  Abstract}
\vskip 8pt\noindent
A magnetic origin for the honeycomb reconstruction of
metallic surfaces with three-fold symmetry  like Pb/Ge (111)
is proposed.  Assuming  that the groundstate is 
an antiferromagnet insulator over
the triangular lattice of adatom sites (Pb), we demonstrate that
the former 
is simultaneously unstable to canting and to a structural
distortion 
if the surface is soft enough. We therefore predict a net magnetization  
over the reconstructed surface at sufficiently low temperature.

\bigskip
\noindent
PACS Indices:  75.30.Pd, 73.20.At, 79.60.Dp 
\vfill\eject

The ($111$) surface of Ge covered  with $1\over 3$ of a monolayer
of Pb (or Sn) adatoms displays a striking reconstruction 
upon cooling.$^{1-3}$  At room temperature the surface is metallic,
with the adatoms forming a triangular lattice.  However, at
low temperatures, $T < T_*\cong 250\,{\rm K}$, a honeycomb relief
forms over the triangular adatom structure.
This  transition from a  
$\sqrt 3\times\sqrt 3$ surface
structure at ambient temperature to a   reconstructed $3\times 3$
surface at low temperature is observed directly via scanning electron
microscopy (STM) and by low-energy electron diffraction
(LEED) techniques.
The  reconstruction is also reflected in the electronic
structure  of the surface.  In particular, both
angle-resolved photo emission spectra (ARPES) 
and electron energy loss spectra (EELS)
find     that  a pseudogap opens 
at the Fermi surface of the surface  conduction band
at temperatures below  $T_*$.$^{4,5}$ 
Such  measurements
suggest that the reconstructed surface has an {\it insulating}
groundstate.

It was originally proposed that the $3\times 3$ surface  reconstruction is
due to a charge-density wave (CDW) instability within the metallic
$\sqrt 3\times\sqrt 3$ surface.$^1$  The absence of good
nesting properties of the Fermi surface at room temperature$^{2-5}$
indicates, however, that this is unlikely.
To account for the phenomenon that remains, we propose
instead that the reconstructed surface is an antiferromagnetic
Mott insulator$^{6,7}$  composed of localized spin-1/2 moments 
associated with the dangling chemical bond left at each adatom site.
Both the narrowness of the surface conduction
band$^{4,8}$ and the presence of the  low-temperature pseudo-gap 
feature$^{1,5}$
make this idea plausible.  In the limit of large on-site repulsion,
we are then left with  an antiferromagnetic
Heisenberg model description for the electronic  spins
that remain over the
sites of the triangular lattice of adatoms. 
A classical analysis of this model in combination with
magnetostriction phenomenology$^9$ leads us to 
conclude  that the magnetic frustration intrinsic to antiferromagnetism
on a   triangular lattice drives a soft enough
$\sqrt 3\times \sqrt 3$
surface to buckle or dilate with the periodicity of 
a honeycomb relief.   We therefore
predict that the low-temperature $3\times 3$ surface reconstruction
has a structural component.  The original LEED evidence
for the reconstruction of the Pb/Ge (111) surface$^1$
as well as recent surface X-ray diffraction
studies$^{10,11}$ support this point of view.
Coincident with the honeycomb distortion is a canted antiferromagnetic
spin arrangement with reduced frustration (see Fig. 1).  We
therefore also predict a net magnetization  over
the  reconstructed surface at low enough temperatures.

We begin the theoretical discussion at zero temperature,
where a Mott insulator state$^{6,7}$ is assumed for the reconstructed
metallic surface.$^1$  Specifically, suppose that a localized
spin $s ={1\over 2}$  moment, $\vec S_i$, lies at each site, $i$, of the
triangular lattice coincident with the Pb (or Sn) adatoms.  The former
represents the spin 
 associated with the 
dangling chemical  bond on the Pb/Ge (111)
surface.  The simplest description of this system
is an antiferromagnetic Heisenberg model
$$H = \sum_{\langle ij\rangle} ( J_{ij} \vec S_i\cdot\vec S_j
           + \delta J_{ij} S_i^z S_j^z )
           - \vec h\cdot\sum_i \vec S_i\eqno (1)$$
summed    over nearest-neighbor bonds $\langle ij\rangle$
on the triangular lattice.  Here, $J_{ij} > 0$ denotes
the exchange coupling constant over such bonds,
while $\vec h$ denotes the external magnetic field.  
Also, the parameter $\delta$ is a measure of the magnetic anisotropy
along the $z$-axis perpendicular to the surface
that originates from the spin-orbit interaction.
We now take the key step in the theory by supposing   that a   honeycomb
lattice distortion accompanies the Mott insulator state (1),
and that this results in two  different exchange coupling constants,
$J$ and $J^{\prime}$, over the undistorted and distorted bonds, respectively
(see Fig. 1).  The buckled (dilated)  lattice is achieved by raising (expanding)
or lowering (contracting) the regions in the
vicinity of the  honeycomb  ($a$ and $b$ sites) with respect 
to those regions in the vicinity of the  centers ($c$ sites) of the honeycomb.
The phenomenology
$$J_{ij} = J_0 + J_1(\eta_i + \eta_j)
\eqno (2)$$
for the exchange coupling constant of the distorted lattice is
compatible with these assignments.  Here $J_1$
is the  first order
magnetostriction coefficient$^{9}$
with respect to the scalar field $\eta_i$, which
can represent either perpendicular displacements, $u_z$,
or lateral dilations, $\partial_x u_x + \partial_y u_y$,
of the (triangular lattice) surface in the vicinity
of adatom $i$.
Equations (1) and (2) represent the essential components of
the theory.

We  now analyze the  isotropic ($\delta =0$)
antiferromagnetic model (1)  in the classical
limit, $s\rightarrow\infty$, in which case the spin operator
$\vec S_i$ reduces to a  3-vector of magnitude $s$.
The ground state for the honeycomb arrangement of exchange coupling
constants $J_{ab} = J$ and $J_{ac} = J^{\prime} = J_{bc}$
indicated in Fig. 1 is a {\it canted}
three-fold symmetric spin configuration
over the three sublattices
$a$, $b$, and $c$, of the honeycomb-distorted  lattice. 
This canted antiferromagnet  has a net magnetization
$$\vec S = {1\over 3} (\vec S_a + \vec S_b + \vec S_c)
 = {1\over 3} (1 - 2\, {\rm cos}\, \theta)\vec S_c \eqno (3)$$ 
aligned parallel to the sublattice magnetization, $\vec S_c$,
of   the $c$ sites, where $\theta$ represents
the half angle in between spins $\vec S_a$ and
$\vec S_b$ (see Fig. 1).  Since this macroscopic moment must be aligned along
the external magnetic field, $\vec h$, we obtain an energy per
site
$$E_{M}/N = s^2 J {\rm cos} (2\theta) + 2 s^2 J^{\prime}{\rm cos}(\pi -\theta)
- E_Z|\vec S|\eqno (4)$$
for the canted antiferromagnet, where $E_Z \propto |\vec h|$
denotes the Zeeman energy splitting.
If we define  $\Delta J = J^{\prime} -J$, then
the minimum  energy occurs at a   half angle $\theta_0$  such that
${\rm cos}\, \theta_0 = 
[s^2 J^{\prime} + {1\over 3}({\rm sgn}\,\Delta J) s E_Z]/2s^2 J$,
with energy
$$E_{M}/N = -s^2 J - 
{[s J^{\prime} + {1\over 3}({\rm sgn}\,\Delta J)  E_Z]^2\over{2 J}}
+{1\over 3}({\rm sgn}\,\Delta J) s E_Z,\eqno (5)$$
and with   net magnetization
$$\vec S = - {1\over 3} \Bigl({\Delta J\over J}
+{{\rm sgn}\,\Delta J\over 3} {E_Z\over{s J}}\Bigr) \vec S_c.\eqno (6)$$
Notice that both an external surface strain [see Eq. (2) and
ref. 9] and an  external magnetic field increases canting, and 
the saturated magnetization, $\vec S$,
as a result.

Above, we have shown how a honeycomb distortion of the frustrated 
antiferromagnet on the triangular lattice 
lowers the magnetic energy via magnetostriction. 
But is such a canted antiferromagnetic state energetically
favorable with respect to    the distorted lattice itself?  
We     return to the magnetostriction
phenomenology (2) 
to address this question. 
Suppose that 
the triangular lattice  
has  a honeycomb distortion,
with lateral  dilations or perpendicular displacements
$\eta_a = \eta_b$ at the honeycomb, and with
lateral dilations or perpendicular displacements
$\eta_c$ at the centers.   
Macroscopic stability of the surface requires that
$$\eta_a + \eta_b + \eta_c = 0.\eqno (7)$$
This fact coupled with expression (5) for the magnetic
energy in the absence of magnetic field,
in addition to the  magnetostriction phenomenology (2), yields
the new expression
$$E_{M}/N = - s^2\Bigl[{3\over 2} J_0 + \bar J_2(\eta_a - \eta_c)^2\Bigr]
\eqno (8)$$
that is valid up to quadratic order in the displacements, where
$$\bar J_2 = {1\over 2} {J_1^2\over J_0}
\eqno (9)$$
is the effective second order magnetostriction coefficient
of the distorted lattice.  Therefore, if the bare energy 
for a lattice
distortion has the effective form
$E_L = {1\over 2} \bar k_0 \sum_{\langle ij\rangle} (\eta_i-\eta_j)^2$,
then the system as a whole, $E = E_L + E_{M}$, is unstable to
a honeycomb-type buckling or dilation    for
soft surfaces such that
$$s^2 \bar J_2 > 2\bar k_0.\eqno (10)$$
Here, $\bar k_0$ denotes 
the relevant effective spring constant for
the case of a buckling distortion.
We therefore observe that the frustrated antiferromagnetism
that exists on the triangular lattice of spins
{\it drives} the honeycomb distortion in order to improve
magnetic energy on  the soft surface (10).  
Last, Eqs. (4)-(6) indicate that this effect is only  {\it enhanced}
by the application of an external magnetic field!

Yet what effect do the  quantum fluctuations that are 
connected with the localized
spin-1/2 moments have on this honeycomb-type lattice 
instability?   It is known that frustration generally
decreases the spin stiffness of a spin-1/2
antiferromagnet.$^{12}$
In our notation, the spin stiffness is essentially the prefactor $s^2$
in expression (8) for the magnetic energy. 
This means that quantum corrections renormalize down $s^2$
to a function $s_R^2$ of frustration that has a minimum
at $J=J^{\prime}$, in which case the frustration is at a 
maximum.   Hence, quantum corrections can only
{\it enhance} the
honeycomb lattice instability (10) driven by frustration
in the antiferromagnet 
(relative to the undistorted triangular lattice [see Eq. (8)]).

We have seen how a canted antiferromagnetic spin arrangement
over the triangular lattice can induce a 
buckling  or dilation instability with a 
periodicity of the honeycomb type.
Strictly speaking, the  effect requires long range order
among      the magnetic  moments.
A ``large-$N$'' analysis of the  
continuum energy functional  
$$E_{M}  = {1\over 2}
\rho_s \int d^2r [(\vec\nabla\vec m)^2 \pm \xi_0^{-2} m_z^2]\eqno (11)$$ 
with respect to the
normalized sublattice magnetization  ($|\vec m| = 1$)
of the  effective {\it classical} Heisenberg model  (1)
indicates that this is indeed the case at temperatures
below a critical temperature,$^{13}$
$k_B T_c \cong 2\pi\rho_s/{\rm ln}|\delta|^{-1}$. 
Here, $\rho_s = s_R^2 J$
denotes the relevant spin stiffness,$^{13}$ 
while $\delta^{\prime} = a^{\prime 2}/\xi_0^2$ is a measure of the
magnetic anisotropy along the $z$-axis perpendicular
to the $3 \times 3$  surface with  lattice
constant $a^{\prime}$.  The anisotropy $\delta^{\prime}$ is assumed to be
small.  A net macroscopic moment
is then  predicted to exist over the ($3\times 3$) surface
at such low temperatures, $T < T_c$.  
At high temperatures,
$T > T_c$, the magnetic correlation length,
$\xi_M(T)$, is finite.  
Clearly, the honeycomb
lattice instability (10) cannot  occur if the magnetic
correlation length is short compared to the lattice constant. 
This suggests the  (implicit) definition $\xi_M(T_*) = a^{\prime}$
for  the paramagnetic
cross-over temperature, $k_B T_* \sim 2\pi\rho_s$.
At  intermediate temperatures 
$T_c < T < T_*$, we expect to have fluctuating
domains of ferrimagnetic order, each with dimensions
of  order $\xi_M(T)$.  A {\it short-range} 
honeycomb lattice instability  (10) is then possible.
In such case, we expect a
diffuse Bragg diffraction  pattern corresponding to the $3\times 3$ 
surface.
 
In the regime of high temperatures, $T > T_*\sim J/k_B$,  our simple
model [Eqs. (1) and (2)] thus predicts a paramagnetic insulator state
over an {\it undistorted} triangular lattice.   This prediction,
clashes, however, with experimental evidence for the existence of  a
metallic surface at ambient temperatures above the reconstruction
temperature, $T_c$.  
The electronic  charge 
fluctuations that are suppressed in the Heisenberg model (1) must
therefore be included.  We resort to the nearest-neighbor Hubbard model
$$H = - \sum_{\langle ij \rangle, \sigma} t_{ij} 
c_{i\sigma}^{\dag} c_{j\sigma} + \sum_i U_i n_{i\uparrow} n_{i\downarrow}
\eqno (12)$$
on the triangular lattice at half filling to address this problem.
Here, $c_{i\sigma}$ denotes the annihilation operator for an electron
of spin $\sigma$ at site $i$, while $n_{i\sigma}$ is the corresponding
occupation number.
In the limit of strong on-site repulsion, 
$U_i\rightarrow\infty$, we recover the
antiferromagnetic insulator state (1) already  discussed,
with exchange coupling constants$^6$
$$J_{ij} = 2 t_{ij}^2(U_i^{-1} + U_j^{-1}).\eqno (13)$$
Let us consider the weak-coupling limit $U_i = 0$ instead,
and suppose that a   honeycomb lattice  distortion
is ``frozen in''.
The electron-phonon interaction then
results in two different  nearest-neighbor hopping amplitudes,
$t_{ab} = t$ and $t_{ac} = t^{\prime} = t_{bc}$,
in between the sublattices 
(see Fig. 1).
Assuming Bloch waves
of momentum $\vec k$ for the one-electron states  over each sublattice,
we obtain a corresponding spectrum $\varepsilon_{\vec k}$
that satisfies the characteristic equation
$$0 = \varepsilon_{\vec k}^3 
- (2t^{\prime 2} + t^2)|\alpha_{\vec k}|^2 \varepsilon_{\vec k}
+t^{\prime 2} t(\alpha_{\vec k}^3 + \alpha_{\vec k}^{*3}),
\eqno (14)$$
where the amplitude
$$\alpha_{\vec k} = e^{i\vec k\cdot\vec a_1} + e^{i\vec k\cdot\vec a_2}
+e^{i\vec k\cdot\vec a_3}\eqno (15)$$
is a  sum over the (overcomplete) basis, $\vec a_1 = a\hat x$ and
$\vec a_{2,3} = (a/2)(-\hat x \pm 3^{1/2}\hat y)$,
of  the triangular lattice (see Fig. 1).
This system has two noteworthy limits: ({\it i}) On the undistorted
triangular lattice, $t = t^{\prime}$,  we have a single band
of surface conduction electrons that disperses as
$$\varepsilon_{\vec k} = 
-t(\alpha_{\vec k} +\alpha_{\vec k}^*).\eqno (16)$$
The   density of states per site at 
the Fermi surface for the half-filled band is 
$N(0)\sim W^{-1}$, where
$W = 9t$ gives the bandwidth.
The Fermi surface itself
at half filling is nearly circular  
and evidently shows no nesting (see Fig. 2, at $T >T_c$).
({\it ii})  In the case of an extreme
honeycomb distortion, 
$t\ll t^{\prime}$ or $t^{\prime}\ll t$,
we have two graphene-type bands
that are separated by a small gap at the vertices ($K_{3\times 3}$)
of the  Brillouin zone boundaries that  correspond to the honeycomb.  
Within this gap
lies     a narrow band,
$$\varepsilon_{\vec k} = \bar t
(\alpha_{\vec k}^3 + \alpha_{\vec k}^{*3})/|\alpha_{\vec k}|^2,\eqno (17)$$
that crosses the Fermi surface at half filling, with an effective
hopping amplitude $\bar t =  {1\over 2} t$ and
$\bar t = t^{\prime 2}/t$ in the respective limits
$t\rightarrow 0$ and $t^{\prime}\rightarrow 0$.  
This band has a global maximum at the center of the Brillouin
zone ($\Gamma_{3\times 3}$) and a global minimum 
at the points  $M_{3\times 3}$
that lie in between neighboring $K_{3\times 3}$ points.
The  Fermi surface at half filling for this low-temperature
($T < T_c$) band structure is displayed in
Fig. 2.  The combined effect of a narrow bandwidth, $\bar W = 8\bar t$, 
and of the ``corners'' present in the 
Fermi surface result in a relatively large
density of states, $N(0)\gsim \bar W^{-1}$,  there.

In light of these observations, we propose the following mechanism
that accounts for the metallicity of the unconstructed
surface.  At high temperatures $T > T_{c}$, the electronic states
at the surface are described by a half-filled nearest-neighbor
Hubbard model (12) over the triangular lattice,
$t = t^{\prime}$, with an on-site
repulsion $U$ below the critical one, $U_c\sim W$, at
the Mott transition.$^{7}$
This is consistent with a metallic state.  At low temperatures,
$T < T_{c}$, the surface buckles or dilates  ($t\neq t^{\prime}$) in
order to lower its magnetic energy ($J\neq J^{\prime}$).
In particular, the honeycomb distortion of the 
triangular lattice of adatoms results in
a narrow band (17) that crosses the Fermi surface
through the effects of the electron-phonon interaction.
The  presence   of an insulating magnetic  groundstate at
zero temperature then requires that the on-site
repulsion $U$  now be {\it greater} than the new critical one,
$U_c\sim \bar W$, at the Mott transition.$^{7}$
This means, in rough terms, that the on-site repulsive energy
must lie inside the window $\bar W < U < W$ of parameter
space. 
Last, the Fermi surface of the reconstructed metallic
surface  (see Fig. 2, at $T < T_c$)
should be unstable to the formation of an SDW
due to the relatively
large density of states
connected  with the narrow low-temperature band (17).$^{14}$  
Since the electron-phonon interaction induces magnetostriction (2) 
through the identity (13), the  previous strong-coupling analysis
indicates that 
this SDW state  is precisely the canted antiferromagnetic
arrangement of spins depicted in  Fig. 1. 
We then  notably predict that the
reconstructed surface has a net magnetization (6)
at low temperatures, $T < T_{c}$.
An analysis of the Hubbard model (12)
within the mean-field approximation 
indicates that such a net moment causes the 
conduction bands (17) of the  reconstructed  surface
to split in a manner similar to Fe.$^{8}$

Yet how does the above proposal compare with experiment?
First,      the presence 
of an SDW  gap
at the Fermi surface
is consistent with evidence that points towards the
existence of an insulating state on 
the  reconstructed Pb/Ge (111) surface at low temperature:
e.g., STM images, 
EELS and surface photo-emission spectra.$^{1, 4}$
Second, the band structure characteristic of a honeycomb distortion
of the triangular lattice  (17) is consistent with
recent ARPES studies of the reconstructed  Pb/Ge (111) surface$^{5}$
that see the gap expected by zone folding at the 
$M_{3 \times 3}$ points of the Brillouin zone. 
Even more striking is the fact that such ARPES experiments
observe iron-type splitting of surface conduction electrons bands!
This of course is consistent with the presence of a net
magnetic moment,$^{8}$ which we propose is due to the
canted antiferromagnetic Mott insulator state 
shown in  Fig. 1.  Last, both LEED and recent X-ray diffraction studies
find evidence for a periodic ($3 \times 3$) structural  distortion
of the reconstructed Pb/Ge (111) surface  that is 
consistent with the  honeycomb distortion
proposed here.$^{1,10,11}$

Contrary to expectations based on the width of
the surface conduction electron band,$^2$ however,
the $3\times 3$ reconstruction presented
by the Sn/Ge (111) surface appears at temperatures below
$T_* \cong 210\,{\rm K}$, which is appreciably 
lower than that corresponding to the
Pb/Ge (111) surface.  And unlike the latter, the Sn/Ge (111) surface
appears to
remains metallic down to a temperature  $T = 110\,{\rm K}$.$^{2,3}$
We speculate that such  discrepancies between the Pb and Sn adatoms
surfaces originate from the elemental
differences in the spin-orbit coupling.
The relatively  weak spin-orbit coupling present in the Sn adatom results
in a smaller magnetic anisotropy  $\delta$ [see Eqs. (1) and (11)],
and thus  in both a lower magnetic rigidity, $k_B T_*$, and 
critical temperature, $T_c$.  

This work
was supported in part by National
Science Foundation grant  No. DMR-9322427
and by  the Spanish grant No. DGES PB95-0202.
The authors are indebted to E.G. Michel and to F. Guinea
for discussions.
 
\vfill\eject
\centerline{\bf References}
\bigskip
\item {1.}  J.M. Carpinelli et al.,
Nature {\bf 381}, 398 (1996). 

\item {2.}  J.M. Carpinelli et al.,
Phys. Rev. Lett. {\bf 79}, 2859 (1997).

\item {3.}  A. Goldoni and S. Modesti, Phys. Rev. Lett. {\bf 79}, 3266 (1997).

\item {4.}  A. Goldoni et al.,
Phys. Rev. B {\bf 55}, 4109 (1997).
 
\item {5.} A. Mascaraque et al.,
 Phys. Rev. B{\bf 57}, 14758 (1998);
A. Mascaraque et al.,
Surf. Sci. {\bf 404}, 742 (1998).
 
\item {6.}  P.W. Anderson, Phys. Rev. {\bf 115}, 2 (1959). 

\item {7.} W.F. Brinkman and T.M. Rice, Phys. Rev. B {\bf 2}, 4302 (1970);
for a review, see A. Georges et al.,
Rev. Mod. Phys. {\bf 68}, 13 (1996).
 
\item {8.} G. Santoro et al.,
Surf. Sci. {\bf 404}, 802 (1998);
S. Scandolo et al.,
Surf.   Sci.    {\bf 404}, 808 (1998).
 
\item {9.}  L.D. Landau and E.M.  Lifshitz,
{\it Electrodynamics of Continuous Media} (Addison-Wesley, Reading, 1960)
chap. 5.

\item {10.} A.P. Baddorf et al.,
Phys. Rev. B {\bf 57}, 4579 (1998).

\item {11.}  A. Mascaraque et al. (unpublished).

\item {12.} J. Bon\v ca et al.,
Phys.    Rev.   B {\bf 50}, 3415 (1994);
J.P. Rodriguez et al.,
Phys.    Rev.   B {\bf 51}, 3616 (1995).

\item {13.} A.M. Polyakov,
{\it Gauge Fields and Strings} (Harwood, New York, 1987).


\item {14.} J.R. Schrieffer et al.,
Phys. Rev. Lett. {\bf 60}, 944 (1988).

\vfill\eject
\centerline{\bf Figure Caption}
\vskip 20pt
\item {Fig. 1}  Shown is the canted antiferromagnetic spin arrangement
over the 
$3 \times 3$ lattice.

\item {Fig. 2}  The Fermi surface (FS) for both the low-temperature
(17) and the high-temperature (16) band structures 
at half filling are displayed.  
Also displayed are
the corresponding  Brillouin zone (BZ) boundaries.

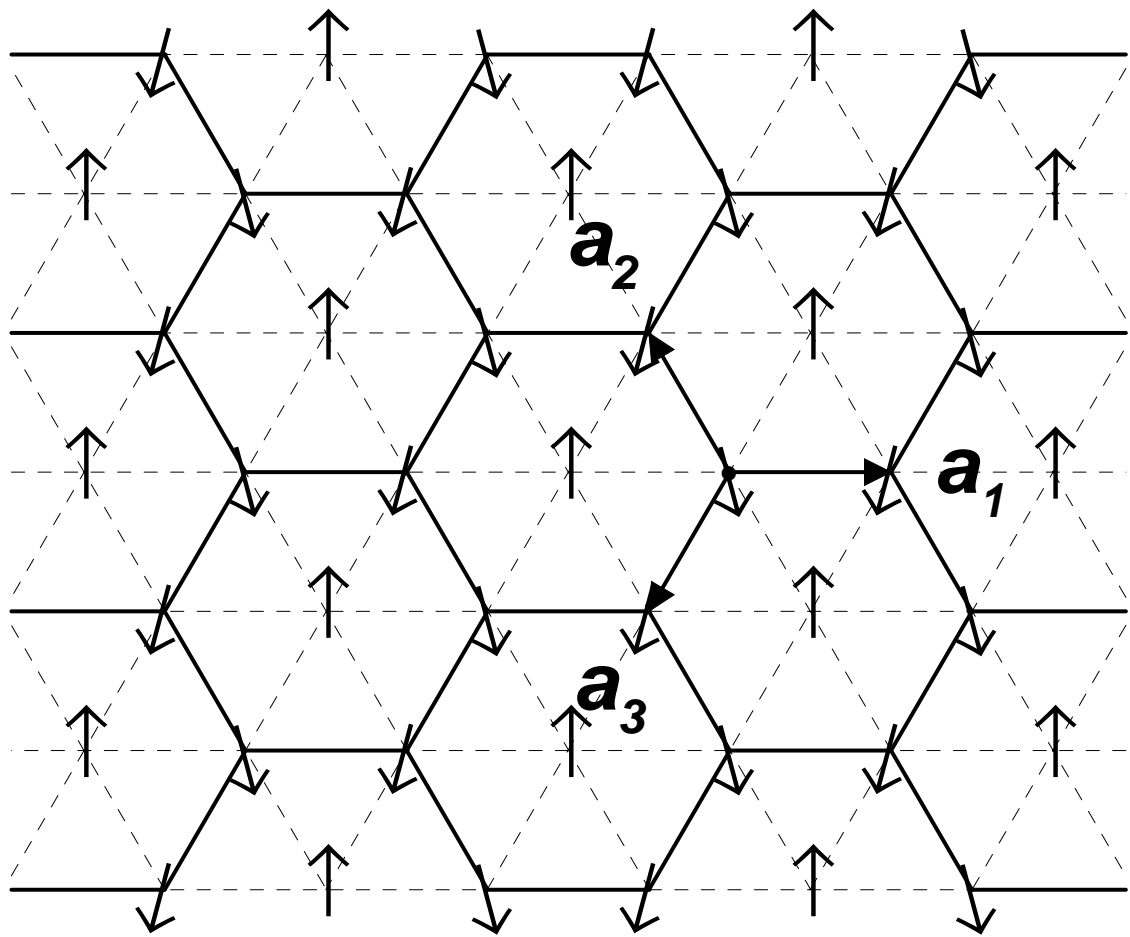
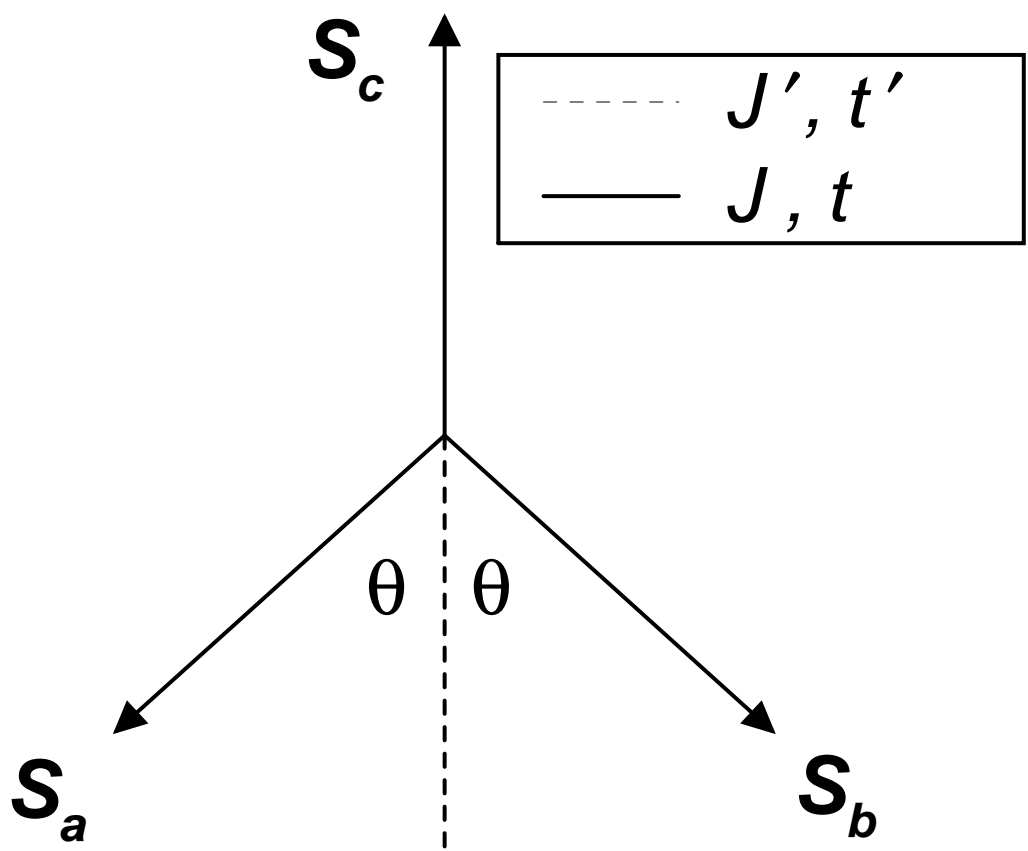
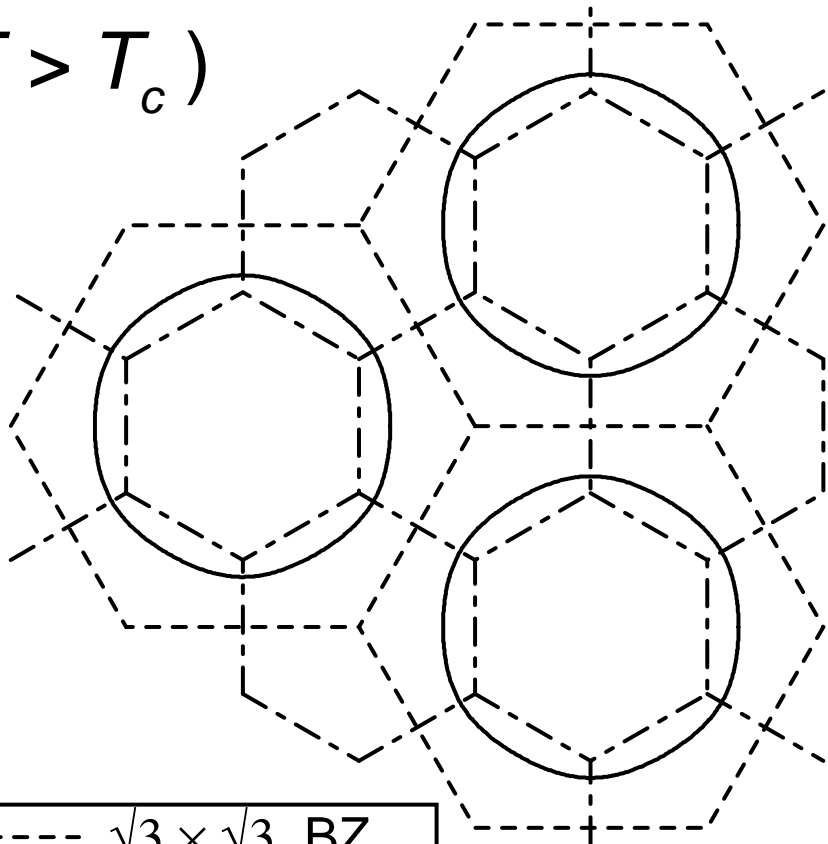
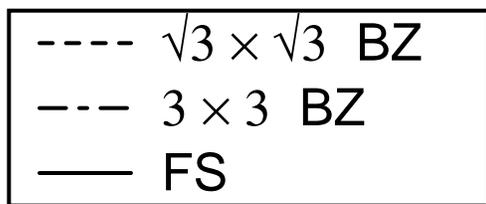
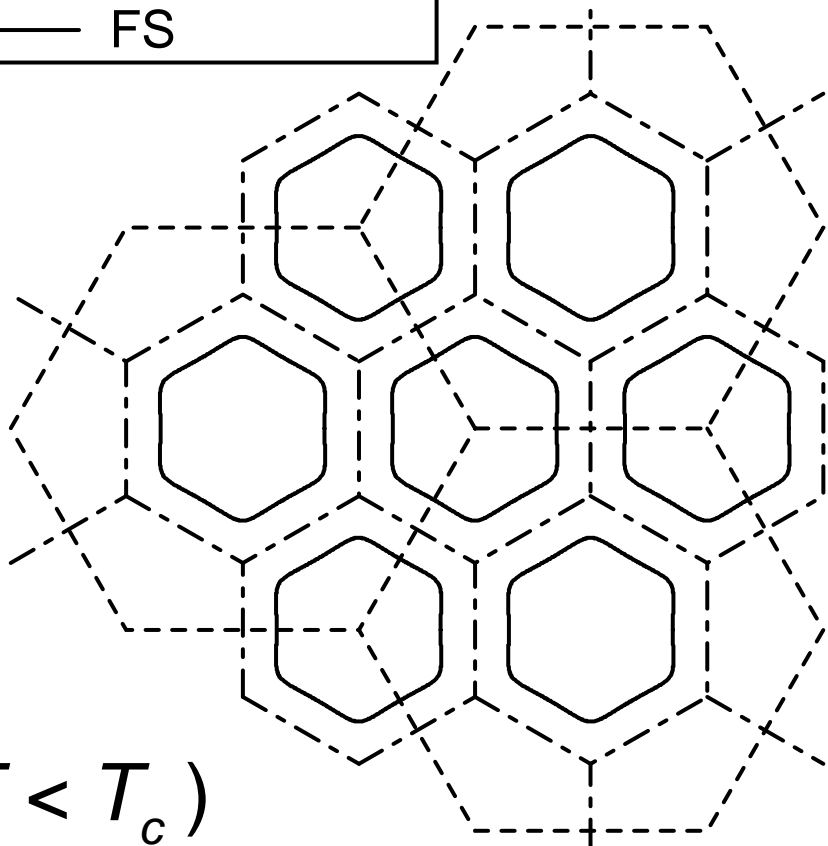
\end